\documentclass[aps,prb,reprint,showpacs,groupedaddress]{revtex4-1}

\usepackage{bm}
\usepackage{wasysym}
\usepackage{amssymb}
\usepackage[cp1251]{inputenc}
\usepackage[dvips]{graphicx}
\usepackage{ulem}
\begin{document}

\title{Undamped relativistic magnetoplasmons in lossy two-dimensional electron systems\\}

\author{V.A. Volkov}
\email{Volkov.V.A@gmail.com}
\affiliation{Kotelnikov Institute of Radio-engineering and Electronics of the RAS, Mokhovaya 11-7, Moscow 125009, Russia\\
}
\affiliation{Moscow Institute of Physics and Technology, Institutskii per. 9, Dolgoprudny, Moscow region 141700, Russia\\}

\author{A.A. Zabolotnykh}
\affiliation{Kotelnikov Institute of Radio-engineering and Electronics of the RAS, Mokhovaya 11-7, Moscow 125009, Russia\\
}
\affiliation{Moscow Institute of Physics and Technology, Institutskii per. 9, Dolgoprudny, Moscow region 141700, Russia\\}

\date{\today}

\begin{abstract}
We address electrodynamic effects in plasma oscillations of a lossy 2D electron system whose dc 2D conductivity $\sigma_0$ is comparable to the speed of light $c$. We argue that the perpendicular constant magnetic field $B$ causes astonishing features of magnetoplasma dynamics. We show that plasmon-polariton spectra can be classified using a “relativistic” phase diagram $\sigma_0/c$ versus $B$. An extraordinarily low damping branch in magnetoplasmon-polariton spectra emerges at two phases of this diagram. Some magnetoplasmons at these phases are predicted to be undamped waves.
\end{abstract}

\pacs{73.21.Fg, 73.20.Mf, 78.67.Wj, 52.27.Lw}
\maketitle

\section{Introduction}
The typical frequencies of plasma waves in three-dimensional (3D) solid-state systems (3D plasmons and surface plasmons) are in the visible and ultraviolet range \cite{Pitarke}. The plasmons have found a number of technological applications in chemical sensing, biomolecular detection, light manipulation, and information processing \cite{Malmqvist,Schuller,plasmonics}. On the other hand, plasmons in 2D electron systems (ESs) have gapless spectra \cite{Stern,Grimes,Allen}, which makes them interesting in the context of the development of novel microwave and terahertz devices \cite{Dyakonov,Kukushkin,Dyer}.
In addition, 2D plasmons have a unique degree of freedom owing to the ability to easily tune their frequency using external electric and magnetic fields. 

2D plasma oscillations are best studied in traditional semiconductor systems such as 2D electron gas in heterostructures GaAs-AlGaAs \cite{Allen}. Nevertheless there is a growing interest in the study of the plasmons in new implementations of 2DESs, including graphene \cite{Fei,Chen,Poumirol,Chak, Horing, Roldan} and topological insulators \cite{Politano}. 
Unexpected manifestations of 2D plasmons are predicted in cosmic dusty plasmas \cite{Hartmann}.

Weakly damped 2D plasmon modes are of particular interest.
The main factor that affects the plasmon damping rate is electron collisions with impurities and phonons. For example, consider a lossy 2D system, where the electron relaxation time $\tau$ and, consequently, the plasmon damping rate Im$\omega_p$ are finite. In the simplest case (without retardation), the complex dispersion relation of 2D plasmons has the form
\begin{equation}
 \label{plasm}	
 \omega_{p}=\omega'+i \omega''=\sqrt{2\pi n e^2q/\varkappa m-1/4\tau^2}-i/2\tau,
\end{equation} 
where $\omega'$, $\omega''$ are real and imaginary parts of $\omega_p$, $n$ is the 2D electron concentration, $\varkappa$ is the background dielectric constant, $m$ is the electron effective mass, and $q$ is the 2D wave vector of the plasmon. Note that at small wave vectors, the plasmons become overdamped, i.e., Re$\omega_{p}= 0$. 

Surprisingly, a weakly damped 2D plasmon mode has been recently observed \cite{Muravev2,Gusikhin} in the regime when ordinary 2D plasmons are overdamped. The authors of papers \cite{Muravev2,Gusikhin} suppose that this puzzling mode is due to electrodynamic effects.

When electrodynamic effects are taken into consideration, plasmons are often called plasmon-polaritons (PPs). 
Consider a 2DES with the dc conductivity $\sigma_0$ in free space (we use Gaussian units, and therefore $\sigma_0$ has velocity dimensionality; for SI units, see Appendix \ref{SI}).
What if the 2D conductivity of the system becomes more than $c$? The authors of papers \cite{Chaplik,Falko} studied the properties of 2D plasmons in the relativistic regime when the dc conductivity $\sigma_0=e^2n\tau/m$ is on the order of $c/2\pi$. It was shown \cite{Falko} that the PPs have low damping, even in a high-collision limit ($\omega'\tau \ll 1$), if $2\pi\sigma_0/c>1$.

In the non-retardation limit, the only kind of plasmon decays slowly even when $\omega'\tau \ll 1$. These are the so-called edge magnetoplasmons \cite{Mast,Glattli,Fetter,Volkov}, which run along the edge of a lossy 2DES placed in a perpendicular magnetic field $\bm{B}$. However, for these plasmons to exist, the magnetic field should be strong enough to satisfy the condition $\omega_c\tau\gg 1$, where $\omega_c=|e|B/mc$ is the electron cyclotron frequency (for theoretical details, see \cite{Volkov1988,Volkov1991}). Recently it has been demonstrated that the edge magnetoplasmon damping in graphene can be lower than that in GaAs systems \cite{Kumada}.

Without a magnetic field, the retardation effects were taken into account theoretically for PPs in a collisionless $(\tau \to \infty)$ 2DES in Ref. \cite{Stern}, a lossy 2DES in Ref. \cite{Falko}, and a lossy bilayer ES in Ref. \cite{Chaplik2015}. For the first time, retardation effects in the spectra of 2D PPs were observed in GaAs/AlGaAs quantum wells with a high electron mobility \cite{Kukushkin1}.

An external magnetic field may modify significantly a microwave response of the system. For example, observations \cite{Zudov2001} of microwave-induced resistance magneto-oscillations in high-mobility 2DESs have attracted a lot of interest; for a review, see \cite{Dmitriev2012}.

If a constant magnetic field is applied perpendicular to the 2DES plane, the dispersion relation of 2D magnetoplasmons in nonretardation and collisionless limits becomes 
$\omega_{mp}=\sqrt{\omega_c^2+2\pi n e^2 q/\varkappa m}$. 
2D magnetoplasmons interact strongly with terahertz waves and microwaves. As a result, an ultrastrong 2D magnetoplasmon--radiation coupling has been observed in the microwave \cite{Muravev2011} and terahertz \cite{Scalari} range. A giant photoresistivity response was observed \cite{Dai,Hatke} under microwave irradiation of a 2DES in the vicinity of the second cyclotron resonance harmonic. This effect has been recently explained \cite{Volkov2014} by an instability of irradiated magnetoplasma.

The retardation effect on magnetoplasmon-polariton spectra was studied by Chiu and Quinn \cite{Chiu}, but only in collisionless 2DESs.
In an external constant magnetic field, the conductivity tensor is of nondiagonal components, and one needs to take into account retardation of Hall currents ($\sim \sigma_{xy}/c$) as well. It results in an interaction of TM (longitudinal) and TE (transverse) waves, and the physics of the plasma waves in lossy 2DES becomes complicated.
This, probably, explains the fact that the results of Ref. \cite{Falko} were not generalized to the lossy 2D magnetoplasma up to now.
Here, we investigate PP spectra in the lossy 2DES placed in the perpendicular classical constant magnetic field.
\begin{figure}
			\includegraphics[width=8.5cm]{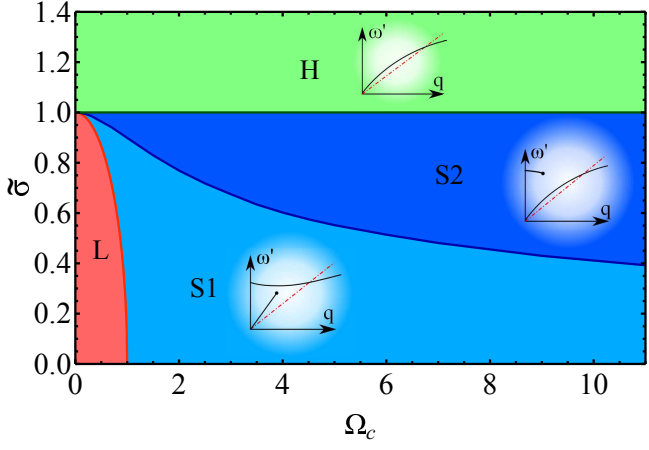}
			\caption{ \label{Fig:Phase}
			Magnetoplasmon-polariton phase diagram in space of dimensionless parameters: 2D dc conductivity $\widetilde\sigma=2\pi \sigma_0/(c\sqrt{\varkappa})$ vs cyclotron frequency $\Omega_c=\omega_c\tau$. 
			The characteristic spectra $\omega(q)$ of high-quality PPs in each phase are presented schematically in the insets (red dash-dotted line corresponds to the 2D light cone); for details, see Figs. \ref{Fig:PP_2}--\ref{Fig:PP_3}.
			In the low-conductivity phase L (light red) and high-conductivity phase H (green) the spectra $\omega(q)$ consist of two branches, but one branch is overdamped. In phases S1 (blue) and S2 (dark blue), an additional extraordinarily low damping branch appears. This branch is gapless in phase S1, gapped in phase S2, and contains a termination point, imaged by a bullet in the insets. The PP damping at this point is zero. The phase boundary L - S1 is defined by the following equation: $\Omega_c^2+\widetilde\sigma^2=1$.
	}
\end{figure}
 
We found that retardation effects in the magnetoplasmon-polariton spectra are controlled not only by the retardation factor ($\sim \sigma_0/c$) but also by the Hall factor ($\sim \omega_c\tau$) with a strong interplay between them.

\begin{figure}
			\includegraphics[width=8.5cm]{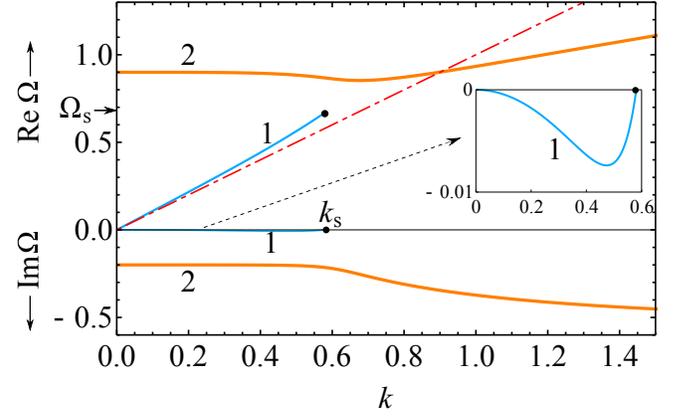}
			\caption{ \label{Fig:PP_2}
			Plasmon-polariton spectra at phase S1 for the following parameters on the phase diagram (Fig. \ref{Fig:Phase}): $\widetilde\sigma=0.8$ and $\Omega_c=0.9$.
			The real, Re$\Omega(k)$, and imaginary, Im$\Omega(k)$, parts of the dimensionless complex PP frequency $\Omega$ as a function of the dimensionless wave vector $k$ are shown, where $\Omega=\omega\tau$ and $k=qc\tau/\sqrt{\varkappa}$. The gapless high-quality branch 1 has a termination point ($k_s$,$\Omega_s$), denoted by the bullet, where Im$\Omega_s=0$. The values $k_s$ and $\Omega_s$ are defined by Eqs.~(\ref{infpoint}). The red dash-dotted line Re$\Omega = k$ corresponds to a 2D light cone. Inset: Enlarged plot of Im$\Omega(k)$ for branch 1.
			}
\end{figure}

\begin{figure}
			\includegraphics[width=8.5cm]{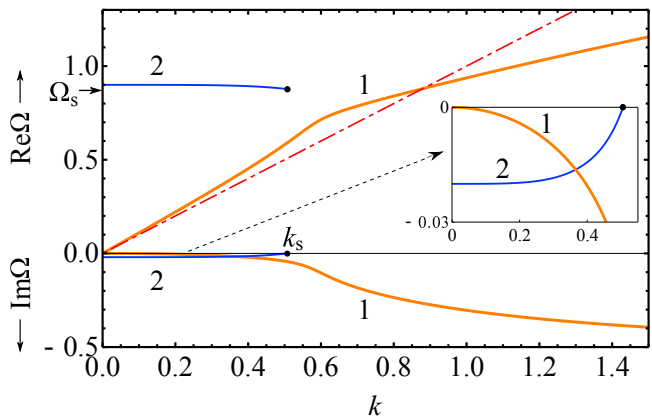}
			\caption{ \label{Fig:PP_2b}
			Plasmon-polariton spectra at phase S2 for the following parameters on the phase diagram (Fig. \ref{Fig:Phase}): $\widetilde\sigma=0.98$ and $\Omega_c=0.9$. The designations are the same as in Fig. 2.
			The gapped extraordinarily low damping branch 2 has a termination point ($k_s$,$\Omega_s$), where Im$\Omega_s=0$ [see Eqs.~(\ref{infpoint})]. Inset: Enlarged plots Im$\Omega(k)$ for branches 1 and 2.
		}
\end{figure}

Let us define the dimensionless cyclotron frequency $\Omega_c=\omega_c\tau$ and the dc conductivity at zero magnetic field $\widetilde\sigma=2\pi\sigma_0/(c\sqrt{\varkappa})$. We represent the results of the above-mentioned interplay on phase diagram $\widetilde\sigma$ vs $\Omega_c$; Fig. \ref{Fig:Phase}. There are four phases. Each phase corresponds to the characteristic type of PP spectra; see Figs.~\ref{Fig:PP_2}--\ref{Fig:PP_3} and insets in Fig. \ref{Fig:Phase}. 

Let us now consider the most interesting phases S1 and S2, in which an extraordinarily low damping PP branch emerges; see Figs. \ref{Fig:PP_2} and \ref{Fig:PP_2b}. S phases are defined by the conditions $\widetilde{\sigma}<1$ and $\widetilde{\sigma}^2+\Omega_c^2>1$; see Appendix~\ref{SI} for these conditions written in SI units.
In these phases the PPs have two branches with $\omega'\neq 0$. At zero wave vector, branch 1 has zero frequency. Another branch 2 has frequency $\Omega_c$. 
One of the branches contains a termination point $(k_s,\Omega_s)$. Damping $\omega''$ at this point equals zero and the PP quality factor $Q\sim \omega'/\omega''$ is infinite; therefore this PP is not damped.
In phases S1 and S2, the point with infinite $Q$ is situated on branches 1 and 2, respectively. The boundary S1-S2 has asymptotic behavior $\widetilde\sigma\to 1-\Omega_c^2/6$ at $\Omega_c\ll 1$ and $\widetilde{\sigma}\propto 1/\sqrt{\Omega_c}$ at $\Omega_c\gg 1$.
 

The absence of the PP damping at the termination point $(k_s,\Omega_s)$ can be qualitatively understood by the following. The PP becomes delocalized in the direction perpendicular to the 2DES plane; i.e., electric and magnetic fields oscillate in space without decreasing. Therefore, PP energy is stored in the whole infinite space. The PP quality factor $Q$ equals $2\pi W/\Delta W$, where $W$ is the energy stored in PP, and $\Delta W$ is the energy loss per period of PP oscillation due to the Joule heat released in the 2DES. $\Delta W$ has finite value due to the finite values of PP fields and conductivity of 2DES, whereas $W$ is infinite as PP is delocalized. That is why $Q$ becomes infinite and $\omega''\sim 1/Q$ equals zero. 

Note that one of the dispersion curves in phases S1, S2, and H has abnormally large derivative $d\omega'/dq>c/\sqrt{\varkappa}$ at small 2D wave vector $q$. This peculiarity should not be surprising since $d\omega'/dq$ does not describe the group velocity as the PPs not only run along the 2DES [in the $(x,y)$ plane] but also move perpendicular to the 2DES (along the $z$-axis). The latter makes a valuable contribution to the group velocity (see Eq. (15) from Ref. \cite{Falko}).

The paper is organized as follows. First, we recall the dispersion equation for 2D magnetoplasmon-polaritons. Second, we analyze the plasmon spectra for different regions in the phase space of dimensionless parameters conductivity $\widetilde\sigma$ vs cyclotron frequency $\Omega_c$. Finally, we draw our conclusions. In Appendix \ref{SI} characteristic values of conductivity, magnetic field, etc., are rewritten in SI units. The Drude model for conductivities is presented in Appendix \ref{Model}.
We determine the field structure and the velocity of the PPs in Appendix \ref{Vel}, and Appendix \ref{Gate} is devoted to PPs in gated 2DESs.

\section{Analysis of dispersion equation} Consider a 2DES positioned in plane $z=0$ and placed in a perpendicular magnetic field $\bf{B}$. One can derive a PP dispersion equation using the Maxwell equations, the continuity equation for the electron charge density $\partial_t\rho+div \textbf{j}=0$, and Ohm's law $\textbf{j}=\hat{\sigma}\textbf{E}$. Here $\rho$ is the charge density, $\bf{j}$ is the electron density current, and $\hat\sigma$ is the 2D conductivity tensor ($\sigma_{xx}=\sigma_{yy}$, $\sigma_{xy}=-\sigma_{yx}$).

We look for solutions in the form $\exp(-i\omega t+i\bf{q}\bf{r})$, where $\textbf{r}=(x,y)$ is the 2D coordinate, $\bf{q}$ is a 2D PP wave vector, $\omega$ is complex frequency, and $\omega''\le 0$. We look for the normal modes that do not grow at $z\to \pm\infty$. The normal modes have the form $\exp(-\beta|z|)$, where $\beta=\sqrt{q^2-\varkappa\omega^2/c^2}$, and the condition Re$\beta\ge 0$ should be held.

Now one can obtain the well-known general dispersion equation for magnetoplasmon-polaritons (see, for example, Eq. (58) from Ref. \cite{Chiu} or Eq. (10) from Ref. \cite{Nakayama}):
\begin{equation}
 \label{pp}	
 \left(1+\frac{2\pi\sigma_{xx}\beta}{-i\omega\varkappa}\right)\left(1-\frac{2\pi\sigma_{xx}i\omega}{c^2\beta}\right)= -\frac{4\pi^2}{c^2\varkappa}\sigma_{xy}^2.
\end{equation} 
If $\bf{B}=0$ ($\sigma_{xy}=0$) Eq. (\ref{pp}) decays into two equations. The first equation describes a TM wave, i.e., wave with two electric field components $(E_x, 0, E_z)$ and a single magnetic component $(0, H_y, 0)$, where we direct the wave vector along the $x$ axis: $\textbf{q} = (q, 0)$. If retardation effects are neglected, the TM wave corresponds to the usual 2D plasmon with spectrum (\ref{plasm}). The second equation describes a TE wave with the field components $(H_x, E_y, H_z)$. The spectra of both waves were analyzed in detail in Ref.~\cite{Falko}. When the magnetic field is nonzero, the TM and TE waves mix to yield new modes.

For simplicity, we use the Drude model for conductivities; see Appendix \ref{Model}.
Then we define the dimensionless magnetoplasmon-polariton frequency and wave vector as follows: $\Omega=\omega\tau$ and $k=qc\tau/\sqrt{\varkappa}$. 
The solutions of the dispersion equation strongly depend on the values of the magnetic field and 2D dc conductivity. This is why we consider four different phases of parameters $\Omega_c$ and $\widetilde\sigma$ (see Fig.~\ref{Fig:Phase}).

\textbf{Phase L} corresponds to the case of low conductivities and weak magnetic fields ($0<\widetilde{\sigma}^2<1-\Omega_c^2$). 
\begin{figure}
			\includegraphics[width=8.5cm]{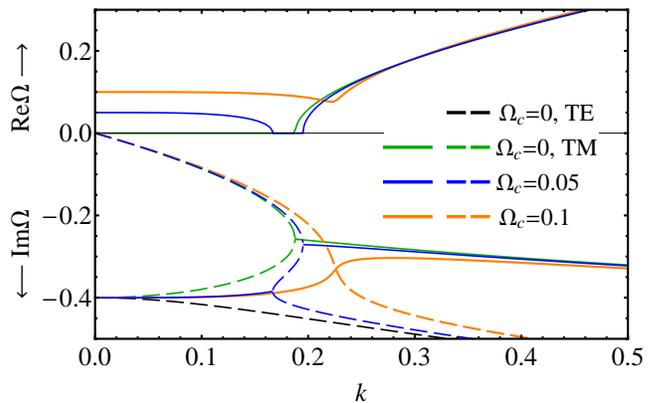}
			\caption{ \label{Fig:PP_1}
			Plasmon-polariton spectra at phase L (see Fig. \ref{Fig:Phase}) at conductivity $\widetilde\sigma=0.6$. The axes are the same as in Fig. \ref{Fig:PP_2}. The dashed lines denote branches with Re$\Omega=0$. The colors correspond to different values of the cyclotron frequency $\Omega_c$: $\Omega_c=0$ (black and green), $\Omega_c=0.05$ (blue), and $\Omega_c=0.1$ (orange).
		}
\end{figure}
In the absence of a magnetic field, we observe TM and TE waves, denoted by green and black lines in Fig.~\ref{Fig:PP_1}. Both waves are overdamped ($\Omega'=0$) at small $k$.
At a nonzero magnetic field, one branch remains overdamped at small $k$ (see dashed blue and dashed orange lines in Fig.~\ref{Fig:PP_1}). The other branch (solid blue and solid orange lines in Fig.~\ref{Fig:PP_1}) has asymptotics at $k\rightarrow 0$ as follows:
\begin{equation}
 \label{acr}	
 \Omega=\Omega_c-i(1-\widetilde\sigma)+\frac{-\widetilde\sigma(\widetilde\sigma-i\Omega_c)}{8\Omega_c(\Omega_c-i(1-\widetilde\sigma))^4}k^4.
 \end{equation}

At $k\rightarrow\infty$, one branch (see dashed lines in Fig.~\ref{Fig:PP_1}) is overdamped $(\Omega\rightarrow -i)$ and
the other branch (solid lines in Fig.~\ref{Fig:PP_1}) has asymptotics of the usual magnetoplasmons: $\Omega\rightarrow \sqrt{\Omega_c^2+\widetilde\sigma k-1/4}-i/2$.

\textbf{Phases S1 and S2} correspond to the case of low 2D conductivity, but a large interval of magnetic fields ($1-\Omega_c^2 < \widetilde\sigma^2 < 1$).
In this regime, an extraordinarily high quality branch appears in PP spectra (branch 1 in Fig.~\ref{Fig:PP_2}). Branch 2 in Fig.~\ref{Fig:PP_2} originates from the usual magnetoplasmon in phase L (see solid orange line in Fig. \ref{Fig:PP_1}). The third branch (it is not denoted in Fig. \ref{Fig:PP_2}) also exists, but it is overdamped at any $k$. This branch has behavior similar to that of the dashed orange line in Fig. \ref{Fig:PP_1}. 

Low-frequency branch 1 has the following asymptotic behavior at $k\rightarrow 0$:
\begin{eqnarray}
 \label{asim}	
 \Omega=k\sqrt{\frac{1}{2}+\frac{1}{2}\frac{\widetilde\sigma^2+\Omega_c^2+1}{\sqrt{(\widetilde\sigma^2+\Omega_c^2+1)^2-4\widetilde\sigma^2}}}-\nonumber\\
 \quad ik^2\widetilde\sigma^2\frac{\widetilde\sigma^2+\Omega_c^2-1}{((\widetilde\sigma^2+\Omega_c^2+1)^2-4\widetilde\sigma^2)^{3/2}}.
 \end{eqnarray}
Branch 1 subsides weakly at small $k$, even if the condition $\Omega'\ll 1$ (i.e., $\omega'\tau\ll 1$) takes place.  High-frequency branch 2 is described at $k\rightarrow 0$ by Eq. (\ref{acr}).

An unexpected result is that in S phases, a point of infinite quality factor $(\Omega''=0)$ emerges in magnetoplasmon-polariton spectra. The extraordinarily high quality branch terminates at this point; the position of this point ($k_s$,$\Omega_s$) is defined as follows:
\begin{equation}
 \label{infpoint}	
 \Omega_s^2=\widetilde\sigma^2+\Omega_c^2-1,  \quad k_s^2=\frac{2(\widetilde\sigma^2+\Omega_c^2-1)}{1+\sqrt{1/(1-\widetilde\sigma^2)}}.
 \end{equation}

This point is situated on branch 1 at phase S1. However, as the magnetic field or conductivity increases, we intersect phase boundary S1-S2, and the termination point moves to branch 2, see Fig. \ref{Fig:PP_2b}.
As $\widetilde\sigma$ tends to unity, $k_s$ tends to zero [see Eq. (\ref{infpoint})], branch 2 in Fig. \ref{Fig:PP_2b} vanishes, and phase S2 transitions to phase H (see Fig.~\ref{Fig:Phase}).

\textbf{Phase H} corresponds to 2DES with high conductivity ($\widetilde\sigma>1$). Note that the asymptotics $\widetilde\sigma \gg 1$ corresponds to the limit considered by Chiu and Quinn \cite{Chiu}.

\begin{figure}
			\includegraphics[width=8.5cm]{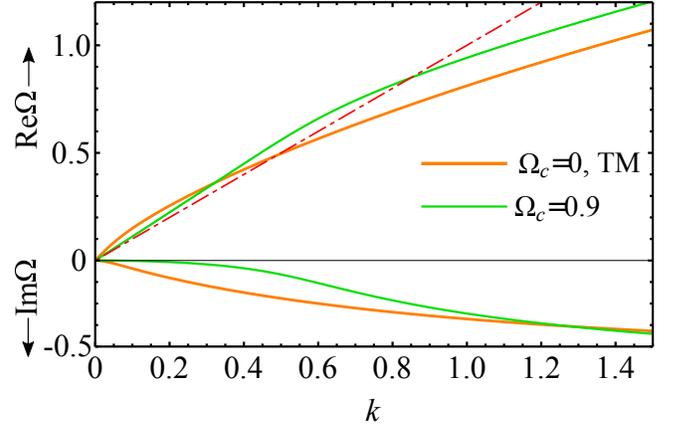}
			\caption{ \label{Fig:PP_3}
			Plasmon-polariton spectra at phase H (see Fig. \ref{Fig:Phase}) at $\widetilde\sigma=1.2$. The designations are the same as in Fig. \ref{Fig:PP_2}. Green and orange colors correspond to different values of the cyclotron frequency, i.e., $\Omega_c=0$ (orange) and $\Omega_c=0.9$ (green). At $\Omega_c=0$, only the TM wave is plotted.
		}
\end{figure}
The PP spectra without/with the magnetic field are presented in Fig. \ref{Fig:PP_3}. There are two branches in phase H. One branch is overdamped at any $k$ (it is not denoted in Fig. \ref{Fig:PP_3}) and has behavior similar to those of the dashed orange line in Fig. \ref{Fig:PP_1}. Another one stems from the branch 1 in Fig. \ref{Fig:PP_2b}. The branch has at $k\rightarrow 0$ asymptotic behavior described by Eq. (\ref{asim}), and a high $Q$ factor ($|\Omega''|\ll\Omega'$). At $k\rightarrow \infty$, this branch behaves like a usual magnetoplasmon: $\Omega\rightarrow \sqrt{\Omega_c^2+\widetilde\sigma k-1/4}-i/2$.

Now let us consider the PP field structure in the $z$ direction.
Note once again that we find solutions that do not increase at $|z|\to\infty$, i.e., Re$\beta=$Re$\sqrt{q^2-\varkappa\omega^2/c^2}\ge 0$. The imaginary part of $\beta$ defines the direction of the PP phase velocity along the $z$ axis. It turns out that all the above considered PPs have phase velocity directed to the 2DES. PPs without a magnetic field \cite{Falko} demonstrate similar behavior. 
Note that one cannot evaluate the group velocity of 2D PPs as $d\omega/dq$ even in dissipationless 2DES. This is because the PP is a nonuniform wave, and it moves along the $z$ direction to the plane $z=0$ from both half spaces $z<0$ and $z>0$.
One should use a more accurate relation (see Eq.~(15) from Ref.~\cite{Falko}) to determine the PP group velocity. In the case of lossy systems, the velocity of signal propagation can be characterized (see, for example, Ref. \cite{Gin}) by the velocity of energy transfer $\textbf{V}=\textbf{N}/w$, where $\textbf{N}=c[\textbf{E}\times\textbf{H}]/4\pi$ is the Poynting vector and $w=(\varkappa \textbf{E}^2+\textbf{H}^2)/8\pi$ is the energy density of the wave. Components $V_x$ and $V_y$ define the signal velocity along 2DES.

The polarization and the velocity of PPs are considered in Appendix \ref{Vel}. Let us discuss the velocity of energy transfer $\textbf{V}$ at the termination point $(k_s,\Omega_s)$. 
It turns out that energy travels from infinity to 2DES and then dissipates in 2DES due to finite value of $\tau$. Moreover, the velocity of energy transfer modulus at the termination point is equal to the speed of light in the medium: $|\textbf{V}| = c/\sqrt{\varkappa}$. 

\section{Discussion and conclusions}

To describe the dynamics of 2DES, we use the simplest classic approach (the Drude model for conductivities; see Appendix \ref{Model}). In general, the Drude model works well in semiconductor systems. For instance, its reliability up to frequencies of several GHz was demonstrated in paper \cite{Burke}. However, the Drude model does not take into account the strong electron correlations, which are responsible for the appearance of shear and magnetoshear modes \cite{Kalman92, Kalman93, BonitzE, BonitzL} in dusty plasmas.

We do not take into consideration quantum effects. That is why, strictly speaking, the conditions $\hbar \omega_c < T$, $\hbar \omega_c \ll E_F$ and $\hbar \omega < E_F$ should be held for our results to be correct (here $E_F$ is the Fermi energy). Nevertheless, the following should be mentioned. In Ref. \cite{Chiu} the magnetoplasmon-polariton spectrum in a clean ($\tau\to \infty$) 2DES was derived in the random phase approximation (RPA) taking into account Landau quantization. In the long-wavelength limit $q\ll R_c^{-1}$ and $q\ll k_F$ (here $R_c=v_F/\omega_c$ is the electron cyclotron radius, $\hbar k_F$ and $v_F$ are the Fermi momentum and velocity), the calculated spectra coincide with those achieved using the Drude model in the collisionless limit. Our PP branch and the point of infinite quality factor are just situated at small wave vector $q$, where the Drude model is applicable; see Appendix \ref{Model}.

In conclusion, we analyzed and classified the spectra of PPs in a lossy 2DES in a transverse classical magnetic field. We showed that the frequency and damping rate of PPs are described by the interplay of three factors: retardation of longitudinal and transverse (Hall) currents, and collisional processes.
The characteristic PP spectra can be classified using a phase diagram in magnetoplasma parameter space, i.e., dimensionless conductivity $\widetilde{\sigma}$ vs dimensionless magnetic field $\Omega_c$.
A relativistic extraordinarily low damping branch of PPs emerged in the two S phases, when the condition $1 - \Omega_c^2 < \widetilde\sigma^2 < 1$ is satisfied. 

This branch is gapless in phase S1 (but has a gap in phase S2) and is described by an unusual dispersion curve that is situated inside the 2D light cone. Nevertheless, the velocity of energy transfer (an analog of group velocity) does not exceed the speed of light. This is due to the fact that the electric and magnetic fields of these PPs oscillate in all three directions, including the direction perpendicular to the 2DES plane. Finally, we demonstrate that the above mentioned branch contains a termination point. The damping at this point is zero and the velocity of the energy transfer modulus is equal to the speed of light.

In experiments \cite{Muravev2,Gusikhin}, a weakly damped PP mode was observed in gated 2DES. In the absence of an external magnetic field, the amplitude of the mode had a sharp peak near $\widetilde\sigma=1$. In the phase diagram (Fig.~\ref{Fig:Phase}), the above region corresponds to a singular point $(0,1)$ where all four phases coalesce. We emphasize that we do not put forward an explanation of the experiments \cite{Muravev2,Gusikhin}. In an infinite gated 2DES, the retardation effects do not influence significantly the magnetoplasmon spectra; namely, additional plasmon branches do not appear (see Appendix \ref{Gate}). Nevertheless, this situation can change in the finite-size 2DES due to the influence of edges of the system. This complicated problem needs special consideration. For example, edge magnetoplasmon-polaritons with the intrinsic long lifetimes can appear in a finite 2DES \cite{Sasaki}.  

\begin{acknowledgments}
We thank I. V. Kukushkin and V. M. Muravev for stimulating discussions and the opportunity to be aware of the results of papers \cite{Muravev2,Gusikhin} before their publication. The work was financially supported by the Russian Science Foundation (Project No. 16-12-10411).
\end{acknowledgments}

\appendix
\section{Conditions for S phases in SI units}
\label{SI}
There are two conditions for the additional branch in PPs spectra to appear (see S1 and S2 phases in Fig. \ref{Fig:Phase}). In Gaussian units they can be written as
$\widetilde{\sigma}=2\pi \sigma_0/c\sqrt{\varkappa}<1$ and $\widetilde{\sigma}^2+(\omega_c \tau)^2>1$,
where $\sigma_0$ is the static conductivity without magnetic field, $c$ is the speed of light (in vacuum), $\varkappa$ is the background dielectric constant, $\omega_c=|e|B/mc$ is the electron cyclotron frequency, $e$ and $m$ are the electron charge and effective mass, and $\tau$ is the electron relaxation time due to collisions with impurities.
Here we discuss these conditions in SI units.

The first condition appears in SI units as
\begin{equation}
	\label{cond1}
	\sigma_0/2\epsilon_0 c\sqrt{\varkappa}<1,
\end{equation}
where $\epsilon_0\approx 8.854 \cdot 10^{-12}$~F/m is the vacuum permittivity. If we assume that the relative dielectric permittivity $\varkappa$ is equal to unity, then one can find using Eq. (\ref{cond1}) that the minimal resistivity $\sigma_0^{-1}$ of the 2DES is 188 ohms$/\square$. Such resistivity corresponds to electron mobility $\mu=16.5$ m$^2$/Vs for electron concentration $2\cdot 10^{15}$ m$^{-2}$ in a typical semiconductor 2DES.

The second condition can be written in SI units as
\begin{equation}
	(\sigma_0/2\epsilon_0 c\sqrt{\varkappa})^2+(\tau eB/m)^2>1.
\end{equation}
This condition is trivially satisfied if $\tau eB/m=B\mu>1$. For typical mobility $\mu=10$ m$^2$/Vs one can find $B>0.1$~T.


\section{Drude model}
\label{Model}

Generally speaking, conductivity $\sigma (\omega,q)$ and dielectric function $\epsilon(\omega,q)$ depend on frequency $\omega$ and wave vector $q$. In 2DESs with infinite relaxation time $\tau$, they can be found, for example, using RPA. 
However, RPA fails to include finite relaxation time effects \cite{Mermin}.

Mermin's approach \cite{Mermin} conserves the local electron number and is widely applied. In the long-wavelength limit ($q \to 0$), one can derive that Mermin's dielectric function becomes $\epsilon(q,\omega)=1- \omega_{p0}^2(q)/\omega^2 (1+i/\omega \tau)$, where $\omega_{p0}(q)$ is the plasma frequency at the dissipationless limit ($\tau \to \infty$). This dielectric function coincides with the usual Drude dielectric function. Therefore, the Drude model is applicable in the long-wavelength limit.

If constant magnetic field is applied perpendicular to the 2DES plane, then the Drude model for conductivities appears as
\begin{equation}
 \label{Drude}	
 \sigma_{xx}=\frac{(1-i\omega\tau)\sigma_0}{(1-i\omega\tau)^2+\omega_c^2\tau^2}, \, \sigma_{xy}=\frac{-\omega_c\tau\sigma_0}{(1-i\omega\tau)^2+\omega_c^2\tau^2}.
\end{equation}
Let us solve the general dispersion equation (\ref{pp}) for plasmon-polaritons using expressions (\ref{Drude}).

We define dimensionless magnetoplasmon-polariton frequency and wave vector as follows: $\Omega=\omega\tau$ and $k=qc\tau/\sqrt{\varkappa}$. 
Using Eqs. (\ref{Drude}), the general dispersion equation for complex frequency $\Omega$ and the real wave vectors $k$ can be written in dimensionless form as  
\begin{equation}
 \label{pp_disp}	
 k^2-2\Omega^2=i\Omega \frac{(1-i\Omega)^2+\Omega_c^2+\widetilde\sigma^2}{(1-i\Omega)\widetilde\sigma}\sqrt{k^2-\Omega^2}.
 \end{equation}
The solutions of Eq. (\ref{pp_disp}), Re$\Omega(k)+i$Im$\Omega(k)$ depend qualitatively on the values of the dc 2D conductivity and the magnetic field. As a result, the PP spectra can be classified using a phase diagram: dimensionless 2D conductivity $\widetilde{\sigma}=2\pi\sigma_0/c\sqrt{\varkappa}$ versus dimensionless cyclotron frequency $\Omega_c=\omega_c\tau$; see Fig.~\ref{Fig:Phase}.

\section{Polarization and velocity of plasmon-polaritons}
\label{Vel}
Let us now consider polarization of PPs. We direct the 2D PP wave vector $\textbf{q}$ along the $x$ axis and define the polarization coefficient $\delta=E_x/E_y$.
One can qualitatively assume that if $|\delta|\ll 1$ then PP is close to the TE wave, and if $|\delta|\gg 1$ then PP is close to the TM wave. The expression for $\delta$ can be written as
 \begin{equation}
 \label{polar}	
 \delta(k,\Omega)=\frac{1-i\Omega}{-\Omega_c}\left(\frac{(k^2-2\Omega^2)((1-i\Omega)^2+\Omega_c^2)}{\Omega^2((1-i\Omega)^2+\Omega_c^2+\widetilde\sigma^2)}+1\right).
 \end{equation}

For example, let us find $\delta$ for the point of cyclotron resonance $k=0$, $\Omega=\Omega_c-i(1-\widetilde\sigma)$ (here we assume $\widetilde\sigma<1$). Using Eq.~(\ref{polar}), one can derive $\delta=-i$, which corresponds to the active circular polarization. 

Let us now find the polarization coefficient for the termination point: $\delta_s=\delta(k_s,\Omega_s)$. Using the definition $\delta_s=|\delta_s|\exp(i\psi)$, one can find
\begin{equation}
  	|\delta_s|=\frac{1-\sqrt{1-\widetilde\sigma^2}}{1+\sqrt{1-\widetilde\sigma^2}},\quad \tan\psi=\frac{\sqrt{\widetilde\sigma^2+\Omega_c^2-1}}{\sqrt{1-\widetilde\sigma^2}},
\end{equation}
i.e., the PP at the termination point $(k_s,\Omega_s)$ has mixed polarization, and the magnetic field modifies only the phase of coefficient $\delta$.

Let us find now the velocity of energy transfer $\textbf{V}=\textbf{N}/w$ at the termination point $(k_s,\Omega_s)$. Here $\textbf{N}=c[\textbf{E}\times\textbf{H}]/4\pi$ is the Poynting vector and $w=(\varkappa \textbf{E}^2+\textbf{H}^2)/8\pi$ is the energy density of the wave.
The averaged in time components of $\textbf{V}(k_s,\Omega_s)$ (at $z>0$ and $\textbf{q}$ along $x$ direction) are

\begin{equation}
\label{pvel}
V_x^{av}=\frac{c}{\sqrt{\varkappa}}\frac{\sqrt{2}\sqrt[4]{1-\widetilde\sigma^2}}{\sqrt{1+\sqrt{1-\widetilde\sigma^2}}},
V_z^{av}=-\frac{c}{\sqrt{\varkappa}}\sqrt{\frac{1-\sqrt{1-\widetilde\sigma^2}}{1+\sqrt{1-\widetilde\sigma^2}}},
\end{equation}
and $V_y^{av}=0$; $V_z^{av}$ changes its sign  at $z<0$. It is clear that energy flow is directed to the 2DES; i.e., energy travels from infinity to 2DES and then dissipates in 2DES due to finite value of 2D conductivity.

\section{Magnetoplasmon-polaritons in gated 2DES}
\label{Gate}

Consider an infinite 2DES in the presence of a metallic gate; $d$ is the distance between the gate and 2DES plane. Usually, $d$ is much less compared to the typical sample size $L$ and the wavelength of light $\lambda$ (we assume that the frequency lies in the microwave or terahertz range). Therefore, $\beta d\ll 1$, where $\beta=\sqrt{q^2-\varkappa \omega^2/c^2}$, as $\beta$ is of the order of $L^{-1}$, $\lambda^{-1}$, or even less. Under the condition $\beta d\ll 1$, the dispersion equation for PP a in gated 2DES appears as follows:
\begin{equation}
	\label{gated_pl}
	\left(1-\frac{4\pi i \omega\sigma_{xx}d}{c^2} \right)\left( 1+\frac{4\pi\sigma_{xx}\beta^2d}{-i\omega\varkappa}\right)= -\frac{(4\pi\sigma_{xy}\beta d)^2}{c^2\varkappa}.
\end{equation}

Using the Drude model (\ref{Drude}) and introducing dimensionless notations $\Omega=\omega\tau$, $k=qc\tau/\sqrt{\varkappa}$, $\Omega_c=\omega_c\tau$, and $\widetilde{\sigma}=2\pi\sigma_0/c\sqrt{\varkappa}$, one can see that dispersion equation (\ref{gated_pl}) is controlled by the parameter $V_{gated}^2\varkappa/c^2$, where $V_{gated}=\sqrt{4\pi e^2 nd/m\varkappa}$ is the velocity of gated plasmons at zero magnetic field. In real 2DES, the above parameter is always less than unity. Consequently, additional plasmon branches and the point of infinite $Q$ factor do not appear; PP spectra are similar to those in the phase L (see Fig.~\ref{Fig:Phase}).

\end{document}